\documentclass[sigconf,nonacm]{acmart}

\usepackage{todonotes}
\usepackage{pifont}
\usepackage{stfloats}
\usepackage{booktabs}
\usepackage{multirow}
\usepackage{makecell}
\usepackage{xcolor}
\usepackage{listings}
\usepackage{lipsum}
\usepackage[inline]{enumitem} 
\usepackage{tabularx}
\usepackage{array}

\newcolumntype{Y}{>{\raggedright\arraybackslash}X}

\renewcommand\footnotetextcopyrightpermission[1]{}

\title{A Progressive Approach to Synthesizable RTL Design Generation Using LLMs} 

\author{Xiangfei Kong}
\affiliation{
  \institution{University of South Florida}
  \city{Tampa}
  \state{Florida}
  \country{USA}
}
\email{xiangfeikong@usf.edu}

\author{Tasnim Tabassum}
\affiliation{
  \institution{University of South Florida}
  \city{Tampa}
  \state{Florida}
  \country{USA}
}
\email{tasnimtabassum@usf.edu}

\author{Marwan Abdelwahab}
\affiliation{
  \institution{University of South Florida}
  \city{Tampa}
  \state{Florida}
  \country{USA}
}
\email{abdelwahab2@usf.edu}

\author{Hao Zheng}
\affiliation{
  \institution{University of South Florida}
  \city{Tampa}
  \state{Florida}
  \country{USA}
}
\email{haozheng@usf.edu}

\begin{document}
\begin{abstract}
Large language models can generate register-transfer-level (RTL) designs directly from natural language specifications. Their failures, however, arise mostly from understanding rather than coding \cite{zhang2026understanding, qiu2025towards}. A specification is informal and ambiguous, the model's interpretation stays implicit, and every misreading is committed silently into Verilog, where only simulation can expose it. Intermediate representations make the interpretation partly explicit, yet existing works don't verify the interpretation against the specification, and repair simulation failures at the code level regardless of where the misreading originated. VeriRefine instead treats specification refinement as a verifiable stage of RTL generation. It progressively refines the prose specification into an explicit, schema-constrained account of design intent, expressed as per-signal Abstract Signal Transition Functions (ASTFs) that commit each signal's logic style, clock domain, and reset behavior before any code exists and ground every behavior in a verbatim specification sentence. The refined specification then passes a five-layer audit spanning soundness, completeness, consistency, FSM integrity, and core RTL design rules, so interpretation errors are repaired at the representation level before any Verilog is generated. Once code is generated, each simulation failure is classified as an understanding error or a coding error and routed back to the corresponding stage for targeted repair. Because every signal's hardware class is fixed during refinement, synthesizability becomes a structural property of the pipeline rather than a post-hoc check. With Claude Sonnet 4.6, VeriRefine reaches 94.0\% functional correctness on RTLLM v2.0 and 98.1\% on VerilogEval-Human v2.

\end{abstract}



\maketitle

\section{Introduction}
\label{sec:intro}


Automating Register Transfer Level (RTL) code generation refers to the use of Large Language Models (LLMs) to translate natural language specifications, hardware descriptions, or functional requirements into synthesizable Verilog code. 
It is a rapidly growing subfield of Machine Learning for Electronic Design Automation (ML4EDA) that aims to streamline chip design, reduce human error, and accelerate the development of digital hardware. Generating RTL from natural language remains challenging because it requires translating ambiguous human specifications into precise, cycle-accurate hardware implementations. Current autoregressive LLMs lack an inherent understanding of hardware concurrency, timing, and structural dependencies, often producing syntactically correct code that contains logic errors or is not synthesizable. Moreover, existing evaluation methods focus largely on syntax and functional simulation, overlooking critical physical design constraints such as timing closure, Power-Performance-Area (PPA) objectives, and hardware security, resulting in designs that may be functionally plausible yet impractical for deployment.

Underlying these issues is a core difficulty: the semantic gap between the input specification and the output RTL code. A natural language specification is informal, ambiguous, and often incomplete, while synthesizable RTL is a formal, cycle-accurate description constrained by concurrency, timing, and area and power budgets imposed by the target technology. The two sit at fundamentally different levels of abstraction, which makes closing the gap between them the central challenge of LLM-based RTL generation.

Most prior work maps natural language specification directly to Verilog code, requiring the model to traverse both levels of abstraction at once \cite{thakur2024verigen, liu2023verilogeval, liu2024rtlcoder}. VeriGen \cite{thakur2024verigen} reports that its failures frequently originate in the specification rather than the model, observing that ambiguous prompts admit multiple valid interpretations, such as when a description does not state whether a reset is synchronous or asynchronous, and that the model frequently omits behavior the prompt leaves implicit. This effect can be quantified on VerilogEval \cite{liu2023verilogeval}, which the same RTL designs are paired with specifications written at different levels of abstraction. \textit{VerilogEval-human} was created by manually converting 156 HDLBits problems, including state-transition diagrams, waveforms, truth tables, and Karnaugh maps, into text-only specifications that emphasize behavioral intent and design functionality. In contrast, \textit{VerilogEval-machine} uses specifications automatically generated from the reference RTL by GPT-3.5, which often expose implementation-level details and closely mirror the underlying code structure. Although both subsets target the same designs, functional pass@1 drops substantially on \textit{VerilogEval-human}. Evaluated with GPT-4, for instance, achieves 60\% pass@1 on the \textit{VerilogEval-machine} but only 43.5\% on the \textit{VerilogEval-human}. The same disparity persists for a model fine-tuned specifically for RTL generation. RTLCoder \cite{liu2024rtlcoder} achieves 62.5\% pass@1 on \textit{VerilogEval-machine}, but only 36.7\% on \textit{VerilogEval-human}. Because the target designs remain unchanged and only the specifications differ, the performance loss cannot be explained by model capacity or design complexity. Rather, it reflects the challenge of recovering implementation-level details from intent-level descriptions, highlighting the semantic gap between natural-language specifications and RTL implementations.

Recent work has begun narrow this gap by inserting a structured intermediate representation (IR) between the specification and the generated code. The study \cite{delorenzo2025abstractions} prompts GPT-4o to generate an IR before producing Verilog, thereby separating the reasoning about circuit logic from the writing of syntactically correct code, and reports an improvement in functional pass@1 to 60.4\% on \textit{VerilogEval-Human v1.0.0}. This confirms that making intermediate structure explicit helps recover the implementation details that an intent-level specification leaves implicit. However, existing approach remain a open-loop, in which the IR is generated once and input directly for RTL generation without validation. As a result, the generated IR may omit behaviors described in the specification, introduce unsupported behaviors that are not grounded in the specification, or make assumptions that are inconsistent with the intended design semantics. Because no mechanism exists to verify specification coverage or representation faithfulness, these errors can propagate directly into the generated RTL. In addition, evaluation is typically limited to functional simulation, leaving synthesizability and implementation quality metrics such as timing and area outside the generation process.





To address above limitations, we propose an agentic framework that treats the intermediate representation as a verifiable design artifact, rather than as the transient, unverified prompting step. Given a specification, an LLM first generates a schema-constrained IR that explicitly captures signal behaviors, interfaces, timing semantics, and architectural intent. The generated representation is then audited for specification coverage and internal consistency to ensure that all required behaviors are represented, every IR entry is grounded in the source specification, and no conflicting assumptions are introduced. Only after the representation passes these validation checks is it used for RTL generation. The resulting RTL is evaluated for functional correctness, synthesizability, and PPA objectives. When violations are detected, feedback is propagated back to the representation level, where the design intent can be refined before regenerating RTL. By localizing errors at the abstraction level rather than the implementation level, the proposed framework enables systematic correction of representation errors before they propagate into the final hardware design. Finally, we evaluate our framework on RTLLM 2.0 and VerilogEval-Human v2 benchmarks.

The main contributions of this work are as follows:

\begin{itemize}
    \item We developed VeriRefine, the first framework to operationalize abstraction refinement in LLM-based RTL generation, which progressively refines an abstract design specification in natural language to RTL Verilog, producing higher functional correctness and synthesizability than direct generation while keeping PPA quality on par with the strongest agentic baseline.
    \item We proposed the Abstract Signal Transition Functions (ASTF), a per-signal intermediate representation that bridges the semantic gap between the natural language specifications and Verilog for functional correctness and synthesizability. 
    \item Experiments on \textit{RTLLM v2.0} and \textit{VerilogEval-Human v2} with Claude Sonnet 4.6 show 94.0\% and 98.1\% functional correctness, improvements of 30.0\% and 12.8\% over direct generation, within two designs of the state-of-the-art VerilogCoder at roughly a quarter of its token cost.


\end{itemize}
The architecture of the framework is described in detail in Section ~\ref{sec:method}.

\section{Background and Related Work}
\label{sec:related}

\subsection{Non-Agentic LLM-Based RTL Generation}
Early and foundational efforts in LLM-driven RTL generation operate without multi-agent coordination, relying instead on supervised training, reinforcement learning, or inference-time prompting to improve generation quality — each making meaningful progress but sharing a structural inability to decompose complex specifications or maintain inspectable design intent across generation stages.
RTLCoder~\cite{liu2024rtlcoder} releases an open-source 7B model matching GPT-3.5 on Verilog generation, but if the model misreads the specification semantically during its single forward pass, there is no correction mechanism — the misinterpretation is silently committed to the generated code. CraftRTL~\cite{liu2025craftrtl} targets non-textual specification artifacts (Karnaugh maps, state diagrams, waveforms) with correct-by-construction synthetic data, yet still treats generation as a single forward pass with no inspectable intermediate stage. CodeV~\cite{zhao2025codev} bootstraps higher-quality training pairs by generating NL summaries \emph{from} Verilog, but the resulting model remains a black box with no way to verify its own functional intent. ScaleRTL~\cite{deng2025scalertl} distills long chain-of-thought traces from DeepSeek-R1 to build the first RTL reasoning model outperforming 18 baselines by 18.4\%, but inherits any errors and biases present in the teacher's reasoning. VeriReason~\cite{wang2025verireason} applies GRPO with testbench feedback to reach 83.1\% pass@5, but remains vulnerable to reward hacking where surface-level patterns satisfy the testbench without capturing true design intent. ChipSeek~\cite{anon2025chipseek} folds EDA synthesis feedback into a PPA-aware hierarchical reward, but computing this reward requires running full synthesis tools on every training sample — a process taking minutes per sample — making the training loop prohibitively expensive at scale. VeriRL~\cite{teng2025veriirl} mitigates reward sparsity via trace-back rescoring, yet still offers no structured representation a reviewer could inspect before RTL is produced. VeriPrefer~\cite{wang2025veriprefer} applies DPO on testbench-derived preference pairs across VerilogEval/RTLLM, but only ranks complete outputs with no visibility into where a failing candidate misinterpreted the specification.
HDLCoRe~\cite{anon2025hdlcore} pairs chain-of-thought templates with RAG over curated RTL snippets to suppress hallucinations, but its static retrieval corpus cannot adapt to specification patterns outside its indexed examples. VeriThoughts~\cite{yubeaton2025verithoughts} is an inference-time prompting approach that structures the LLM's reasoning as an explicit circuit-intent description which is then checked by a formal property verifier, with failures fed back as corrective prompts, yet it can only catch errors that the solver is strictly restricted to checking functional equivalence, it cannot detect many other types of errors or design flaws that cannot be easily expressed as a simple logic-matching property. EvolVE~\cite{anon2026evolve} takes an evolutionary search approach at test time: it generates a population of candidate Verilog implementations, then iteratively mutates and selects survivors based on simulation scores across generations, but this means every candidate at every generation requires a full simulation run, causing cost to explode rapidly as design complexity grows. DeepV~\cite{ibnat2025deepv} grounds generation in a curated retrieval-augmented Verilog knowledge base, yet gains degrade sharply for novel module types outside its corpus. Overall, these non-agentic methods remain bounded by the reasoning capacity of a single model pass and cannot decompose hierarchical specifications or maintain inspectable design state across generation stages.

\subsection{Agentic Multi-Agent Frameworks}

VerilogCoder~\cite{ho2025verilogcoder} combines planning, coding, and AST-based waveform  tracing agents to reach 94.2\% pass@1, but its feedback only reports output-level mismatches rather than diagnosing which internal signal or state diverged from the intended behavior. MAGE~\cite{zhao2025mage} introduces high-temperature sampling with checkpoint-guided debugging as the first open-source multi-agent RTL system, yet its state checkpoints expose only output-level signal values at mismatch points rather than the implementation class, clock domain, or guarded-command logic of each driven signal. VERIMOA~\cite{ping2025verimoa} mixes Base/C++/Python specialist agents with a quality-scored cache, but coordinating three heterogeneous agent types adds substantial inference cost and latency. Spec2RTL-Agent~\cite{yu2025spec2rtl} converts specifications into structured plans via intermediate C++/HLS, cutting human intervention by 75\%, but constrains designs to those that map cleanly onto HLS semantics, excluding many practical RTL patterns. VeriGraphi~\cite{anon2026verigraphi} builds a spec-anchored knowledge graph encoding hierarchy and dependencies before code emission, yet extracting that graph from unstructured specification text is itself an unvalidated, error-prone step with no formal correctness guarantee. VeriOpt~\cite{anon2025veriopt} coordinates Planner/Programmer/Reviewer/Evaluator agents for PPA-aware iteration, but repeated EDA tool calls at each iteration make the pipeline far slower than single-pass methods. Despite these advances, agentic systems still pass context informally between roles as unstructured natural language, risking loss of structural intent across turns --- motivating an explicit, inspectable intermediate artifact rather than implicit agent memory. VeriRefine is most closely related to and inspired by VerilogCoder and MAGE: where VerilogCoder structures the coding workflow around a Task and Circuit Relation Graph that retrieves signal and transition facts per subtask, VeriRefine instead validates a formal signal-level representation before any code is written; and where MAGE routes debug feedback from simulation output directly back to RTL repair, VeriRefine classifies each failure as either a representation error or a code error and routes it to the appropriate abstraction level --- enabling targeted correction that neither system can achieve through post-generation repair alone.

\subsection{Abstraction-of-Thought for Verilog Generation}

Hong et al.~\cite{hong2024aot} showed LLMs reason more reliably across explicit abstraction levels before committing to output, but this general-purpose formulation was not designed for hardware-specific syntax and timing constraints. ComplexVCoder~\cite{anon2025complexvcoder} introduces the General Intermediate Representation (GIR), improving over CodeV and RTLCoder by 14.6\% and 22.2\%, yet its schema is fixed and hand-designed rather than learned. QiMeng-CRUX~\cite{huang2026qimengcrux} learns a structured CRUX space (interface, functions, considerations) transferable to external models, but its three fixed categories may not capture unconventional control structures. Paradigm-Based HDL~\cite{sun2025paradigmbased} routes specifications through circuit-class-specific paradigm blocks, but misclassifying novel or hybrid circuits misroutes the entire pipeline. AutoFSM~\cite{luo2025autofsm} compiles a JSON intermediate deterministically to Verilog, eliminating syntax errors, but is narrowly scoped to FSM-style designs. LLM-FSM~\cite{wu2026llmfsm} uses YAML as a ground-truth format for correct-by-construction synthesis, but remains similarly confined to FSMs. RTL++~\cite{akyash2025rtlpp} encodes Verilog into CFG/DFG graphs enriching structural context, yet requires existing reference RTL, limiting it to refinement rather than blank-page generation. SecFSM~\cite{hu2025secfsm} extends graph IRs with a security knowledge graph for secure FSM generation, but its coverage is bounded by the curated CWE patterns it encodes. AoT-RTL~\cite{delorenzo2025abstractions} is the most direct hardware instantiation of AoT, outperforming Tree-of-Thought on GPT-4o, yet its evaluation stops at module-level benchmarks without demonstrating multi-modular scaling. Lorecast~\cite{anon2025lorecast} inserts pseudocode as an implicit CoT step before RTL emission, but offers no external checkpoint for a verifier to intervene. Veritas~\cite{basuroy2025veritas} compiles LLM-generated CNF clauses deterministically for correctness by construction, but CNF can only express a narrow slice of practical sequential RTL. Crucially, NL2Silicon~\cite{fu2026nl2silicon} shows across 3 frontier LLMs and 6 abstraction formats that representation selection---not model selection---dominates success (3\%--88\% pass-rate spread), yet stops short of proposing one format that generalizes across task categories. Across this literature, every abstraction schema is hand-crafted for a narrow design class, and none couples a learned, generalizable intermediate representation with fine-grained internal-signal feedback for targeted hierarchical repair---the gap our framework closes.
\begin{figure*}[!t]
    \centering
    \includegraphics[width=\textwidth]{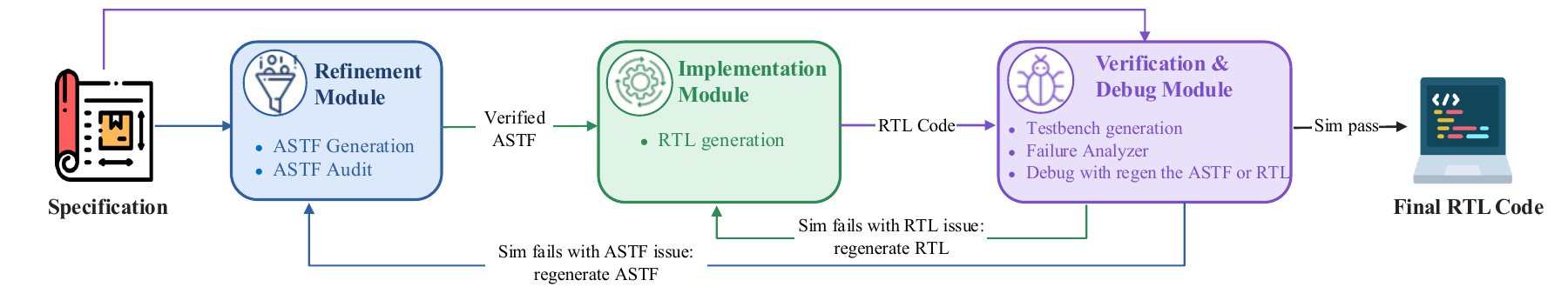} 
    \caption{Overview of the VeriRefine framework}
    \label{fig:workflow}
\end{figure*}


\section{VeriRefine Framework}
\label{sec:method}
VeriRefine is an agentic framework for RTL generation that uses the Abstract Signal Transition Functions (ASTF) to refine structured behavioral information from natural language specifications, enabling functionally correct and synthesizable Verilog to be produced. By separating the refinement step from code generation, VeriRefine enables the extracted representation to be verified independently before RTL is generated, and allows simulation failures to be classified by root cause and routed back to the appropriate stage for targeted correction.
This section describes the overall system architecture (Section \ref{subsec:archi}), the Abstract Signal Transition Functions (Section \ref{subsec:astf}), refinement module (Section~\ref{subsec: refinement module}), implementation module (Section~\ref{subsec:implementation})
, and verification \& debug module (Section~\ref{subsec:debug}).

\subsection{Overview Framework Architecture}
\label{subsec:archi}
VeriRefine organizes RTL generation into three sequential modules coordinated through the ASTF, as illustrated in Figure ~\ref{fig:workflow}. \textit{Refinement Module} takes the natural language specification as input and produces a verified ASTF that explicitly captures the behavioral and structural properties of each signal. The \textit{Implementation Module} then consumes this verified ASTF to generate synthesizable Verilog. The \textit{Verification \& Debug Module} receives the generated Verilog, and a testbench independently generated from the input specification, then runs functional simulation. If simulation passes, the system accepts the Verilog as the final output. Otherwise, the system routes the failure back to the appropriate module for correction, either regenerating the ASTF or the RTL, until the design passes or the iteration limit is reached.


\subsection{Abstract Signal Transition Functions (ASTF)}
\label{subsec:astf}

When an LLM generates Verilog directly from a natural language specification, it must comprehend the design intent and translate it into hardware implementation at the same time, leaving no opportunity to catch misunderstandings before code is produced. A misread timing constraint or a fabricated behavior can result in Verilog that is syntactically valid yet functionally incorrect, and such errors only become apparent during simulation. In practice, simulation is one of the most time-consuming steps in hardware design, and catching an error only at this stage means the entire generation cycle must restart from the beginning.

VeriRefine addresses this through the Abstract Signal Transition Function (ASTF), which makes explicit two types of information that a natural language specification encodes in prose but never separates for independent verification. Formally, for each signal $s$ driven in a design with specification $S$, the ASTF is defined as

$$\text{ASTF}(s) = \bigl(\mathcal{H}(s),\ \mathcal{G}(s)\bigr)$$

$\mathcal{H}(s)$ captures the implementation class of the signal. It records the signal's logic style, its clock domain when the signal is sequential, and its reset behavior when one applies. This information alone is sufficient for the code implementation to fix the always block header for $s$ without consulting any other signal.

$\mathcal{G}(s) = {(c_i, a_i, \sigma_i)}_{i=1}^{n}$ is a priority ordered list 
of guarded actions. 
A plain unordered list would be ambiguous whenever two guards hold in the same cycle, for example a reset asserted while a valid input arrives. Priority ordering resolves every such overlap deterministically, without forcing each guard to be written as a mutually exclusive compound condition that is harder to generate and to audit. 
Each triple pairs a condition $c_i$ with an action $a_i$ and a provenance $\sigma_i$. The condition $c_i$ is a boolean guard written only over signals already declared elsewhere in the ASTF, never over undeclared or external state. The action $a_i$ assigns $s$ whenever $c_i$ holds, and the provenance $\sigma_i \subseteq S$ is one or more verbatim sentences copied character by character from the specification that jointly justify the pair $(c_i, a_i)$. The ordering of $\mathcal{G}(s)$ carries semantic weight rather than serving as mere bookkeeping. Evaluation proceeds from $i = 1$ to $n$, and the first condition $c_i$ that is satisfied determines the value of $s$, the same first match priority order that the emitted RTL realizes as a chain of if and else if statements.

Together, $\mathcal{H}(s)$ and $\mathcal{G}(s)$ fully characterize the behavioral logic of $s$ and leave nothing to be inferred when Verilog is generated. The only state shared across signals is a design level finite state descriptor used by FSM designs, referenced by name inside individual guards but defined once outside any single signal's ASTF.

\begin{figure*}
    \centering
    \includegraphics[width=\linewidth]{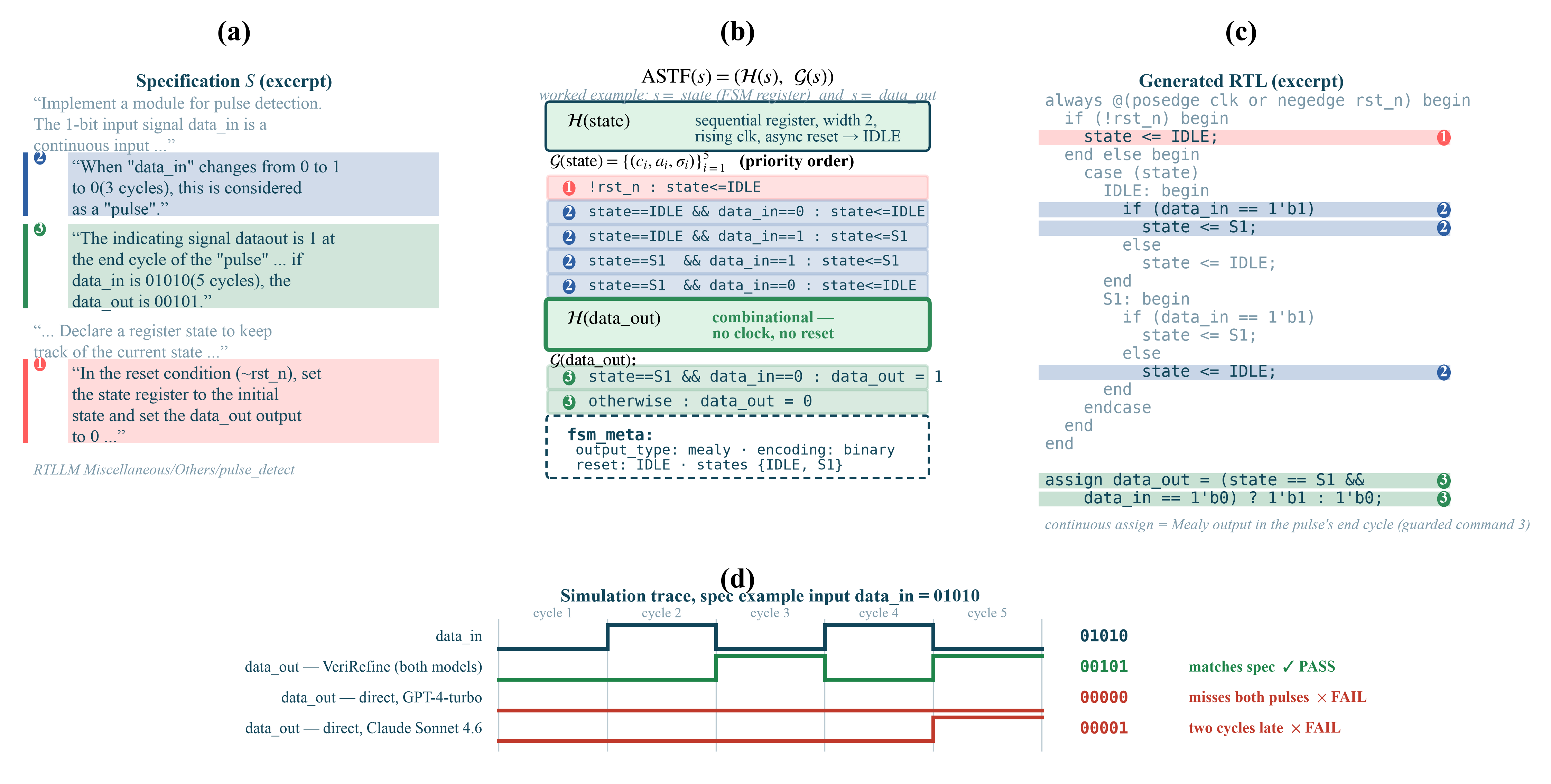} 
    \caption{ASTF traces the \texttt{pulse\_detect} FSM from
    specification (a) through the per-signal records and FSM metadata (b) to the generated RTL (c). Panel (d) simulates the specification's own example input as evidence. Both direct-generation outputs miss the required end cycle timing and fail, while the RTL generated from the ASTF matches the expected trace and passes on the first attempt.}
    \label{fig:astf-example}
\end{figure*}

To illustrate, consider the \textit{RTLLM v2.0} design \texttt{pulse\_detect}, which must assert a one-cycle indicator \texttt{data\_out} in the final cycle of every $0\!\to\!1\!\to\!0$ pulse on \texttt{data\_in}. The specification fixes the timing with an example trace: \texttt{data\_in} 01010 yields \texttt{data\_out} 00101. Direct generation fails this design with both evaluated models, and through the same misreading. Both generated FSMs assert \texttt{data\_out} after the pulse's end cycle rather than in it, one by registering the output inside the clocked always block and one by tying it to the state entered after the pulse completes.

The ASTF settles this timing decision before any Verilog generated. $\mathcal{H}(\texttt{state})$ records a two-bit sequential register clocked on the rising edge of \texttt{clk} and asynchronously reset to \texttt{IDLE}, and $\mathcal{G}(\texttt{state})$ lists the reset command followed by the four state transitions, each citing the sentence that defines the pulse. $\mathcal{H}(\texttt{data\_out})$, in contrast, classifies the indicator as combinational, and the design-level FSM metadata records a Mealy output, because the cited sentence and its example trace require \texttt{data\_out} to assert in the same cycle in which the closing 0 arrives. The Implementation Module of VeriRefine therefore emits the state register as a clocked case statement and \texttt{data\_out} as a continuous assignment guarded by \texttt{state == S1 \&\& data\_in == 0}, and the design passes simulation with both models on the first attempt. Figure~\ref{fig:astf-example} traces this instantiation from the highlighted specification sentences in panel (a), through $\mathcal{H}$, $\mathcal{G}$, and the FSM metadata in panel (b), to the generated RTL in panel (c). Panel (d) then contrasts the simulated traces on this example input, where both direct-generation outputs miss the end-cycle timing and the ASTF-generated RTL matches it.

$\mathcal{H}(s)$ and $\mathcal{G}(s)$ address different failure modes that commonly arise in direct generation. $\mathcal{H}(s)$ enforces synthesizability by committing each signal to a concrete logic style, clock domain, and reset behavior before any code is generated, eliminating the class of errors in which combinational logic is placed inside a clocked always block or reset conditions are left unspecified. $\mathcal{G}(s)$ enforces functional correctness and suppresses hallucinated behaviors by grounding every condition action pair in a verbatim sentence from the specification, ensuring that the representation contains no behavior the specification did not require and reduceing the risk of omitting behavior it did. Because each signal is an independent record, errors in one signal do not propagate to others, and each record can be verified in isolation before simulation is invoked.

\subsection{Refinement Module}
\label{subsec: refinement module}

We design the Refinement Module to derive a verified ASTF from the input natural language specification before any RTL is generated. This stage addresses a central source of failure in LLM-based RTL generation. Many errors arise because the model misinterprets the specification, and sunch errors surface only during the simulation, after a full generation cycle has already been spent. Instead of asking the LLM to translate the specification into RTL directly, we first require it to expose its understanding of the design intent as an auditable signal-level refinement.

\begin{figure}[!t]
    \centering
    \includegraphics[width=\linewidth]{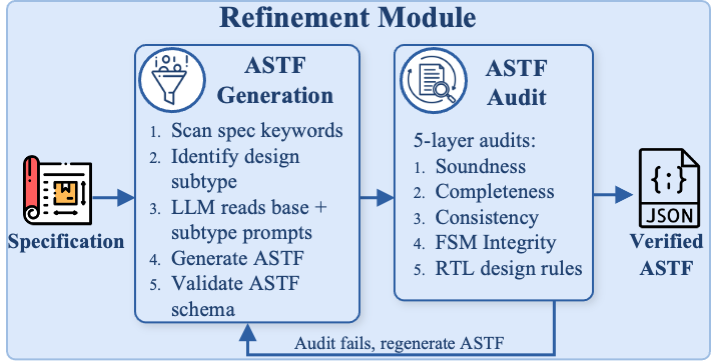} 
    \caption{Refinement Module 
    }
    \label{fig:refinement}
\end{figure}

As shown in Figure~\ref{fig:refinement}, the module reaches a verified ASTF through two quality gates. ASTF generation (Section ~\ref{subsec:astf gen}) prompts the LLM to emit a candidate ASTF as a JSON document. A closed schema (Section~\ref{subsec: schema}) checks the form of this document, rejecting any
output that violates the schema without an LLM call. A five-layer audit (Section~\ref{subsec:audit}) then checks its content, and each violation returns to generation as targeted repair feedback. Form errors are therefore caught mechanically, content errors are caught before RTL generation, and only behavioral errors remain for simulation to expose.
\subsubsection{ASTF Generation}
\label{subsec:astf gen}
Given an input specification, we generate the ASTF in three steps, classification, prompt assembly, and generation that must conform to the ASTF schema (Section~\ref{subsec: schema}).

Classification is a keyword scan over the specification text. Each design type owns a hand-curated list of indicative phrases, such as \textit{finite state} and \textit{Mealy} for FSMs, or \textit{write clock} and \textit{clock domain crossing} for dual-clock designs,  and the scan matches these phrases case-insensitively against the text. The primary type is then assigned in a fixed order from most specific to most general, FSM, dual-clock, sequential, and combinational as the default, because a more specific design usually also contains the vocabulary of the general ones. The \texttt{pulse\_detect} specification of
Figure~\ref{fig:astf-example}, for example, mentions its clock and reset alongside its state transitions, so checking FSM first keeps it from being classified as plain sequential logic. The same scan detects any number of structural sub-types from a catalog of fifteen patterns covering FSMs, counter and boundary behavior, stacks, sequential and combinational dividers, floating-point arithmetic, frequency dividers, pipelines, FIFOs, shifters, registers, edge detection, bit-level operations, table-defined combinational logic, and FSMs with timed datapaths. Sub-types are not exclusive, and detecting one can activate a companion pattern. A FIFO, for example, also activates the boundary pattern, because pointer wrap-around is where FIFO behavior is easiest to misread.

The classification then decides what the generation prompt contains, assembled from four ingredients. The first is a base section shared by all designs. It fixes the output discipline, instructing the model to produce the interface first, then classify each driven signal's logic style, and only then write the guarded commands, and it requires a verbatim specification quotation for every command. The second ingredient is one rule section per detected sub-type. Each section states the structural triggers of its pattern and prescribes the ASTF structure they demand. The FSM section, for example, tells the model when the present state arrives as an input port and an internal state register must therefore not be invented. The third ingredient is one or two worked examples selected by the primary type, each a complete specification-to-ASTF pair. FSM designs, for example, receive one Moore and one Mealy example. The fourth ingredient is optional. When the specification defines behavior through a machine-parseable table, such as a Karnaugh map, a mux-selection table, or a decimal truth table, we parse the exact table cells and inject them into the prompt as authoritative facts, so the model does not re-derive the table from prose. The same parser backs the consistency audit of Section~3.3.3, so the facts the model is given can never contradict the facts it is later checked against.

\subsubsection{ASTF Schema}
\label{subsec: schema}
ASTF generation emits the ASTF as a JSON document, and the formal definition of Section~\ref{subsec:astf} states what that document must contain, but free-form LLM output cannot be checked against a mathematical definition. We therefore constrain every emitted ASTF with a closed JSON schema. The schema forbids any field it does not declare, so a standard validator deterministically rejects and ASTF that omits required information or invents structure the downstream modules cannot consume. The schema is designed around two obligations. It must be able to express every behavior the specification describes, and it must expose that behavior in a form the Implementation Module can translate directly into synthesizable Verilog. 

A document describes one design and contains the module name, optional Verilog parameters, and three signal arrays holding the input, output, and internal signals. All three arrays are always present, though they may be empty. An input signal carries only a declaration record with its name, direction, data type, and bit width, since it is driven externally and assigns nothing. Every output and internal signal is a driven signal and must additionally carry a \texttt{hardware\_type} record serializing $\mathcal{H}(s)$ and a \texttt{guarded\_commands} record serializing $\mathcal{G}(s)$, as the \texttt{data\_out} record of Figure~\ref{fig:astf-record} illustrates, showing the serialized counterpart of the $\mathcal{H}$/$\mathcal{G}$ view in Figure~\ref{fig:astf-example}(b). Bit widths may be integer literals or parameter expressions such as \texttt{WIDTH-1}, so the
same schema covers fixed-width and parameterized designs.

The \texttt{hardware\_type} record fixes the three decisions that determine RTL structure. The \texttt{logic\_style} field selects a combinational, sequential, or latch implementation. The \texttt{clock\_domain} field names the clock and active edge for flip-flops, or the enable level for latches, and accepts a list of domains for multi-clock designs such as asynchronous FIFOs. The \texttt{reset\_behavior} field names the reset signal, the reset value, and whether the reset is synchronous or asynchronous.
Conditional constraints tie the three fields together, turning the applicability conditions stated informally in Section~\ref{subsec:astf} into machine-checkable rules. A combinational signal must leave clock and reset null, and a sequential or latch signal must name a clock
domain, so a contradictory classification is rejected before any Verilog is generated.

The \texttt{guarded\_commands} record lists the
$(c_i, a_i, \sigma_i)$ triples of $\mathcal{G}(s)$. Each entry stores the condition and action as one \texttt{gca\_expression}
string in the fixed form \texttt{guard : action}, and stores the
provenance $\sigma_i$ in a \texttt{spec\_citation} field that must quote the specification verbatim, a requirement the soundness audit of Section~\ref{subsec:audit} enforces. Array order is the priority order defined in Section~\ref{subsec:astf}. For logic that repeats over the elements of a vector, such as a per-bit permutation, a parametric \texttt{for\_commands} entry instead gives an index variable, an inclusive index range, and a \texttt{gca\_expression} over that index. Each such entry maps to one generate-for loop for combinational signals or one for loop inside an always block for sequential signals.

Two design-level constructs sit outside the per-signal records. An FSM design provides a state transition table whose rows map a
(from-state, input-condition) pair to a (to-state, output) pair,
and a non-empty table obligates an \texttt{fsm\_meta} record naming the state register, the encoding (binary, one-hot, or Gray), the Moore, Mealy, or mixed output type, the reset state, and the complete state list, as the bottom of Figure~\ref{fig:astf-record} shows. A design with counters or accumulators may add \texttt{boundary\_conditions} entries stating the exact behavior of a signal at each minimum or maximum limit, which guards against off-by-one errors at wrap-around.

Schema validation is the final step of ASTF generation
(Figure~\ref{fig:refinement}) and divides error detection
with the audit that follows (Section~\ref{subsec:audit}). The validator catches structural errors, such as a driven signal without a guarded command or a sequential signal without a clock, mechanically and without an LLM call. The four audit layers then examine only well-formed ASTFs for semantic errors such as unsupported or missing behavior. This division keeps the audit focused on meaning and gives the Implementation Module a hard structural guarantee about its input.

\begin{figure}[t]
\begin{lstlisting}[
  basicstyle=\scriptsize\ttfamily,
  columns=fullflexible,
  keepspaces=true,
  breaklines=true,
  breakindent=1.5em,
  postbreak=\mbox{\(\hookrightarrow\)\ },
  frame=single,
  framerule=0.3pt,
  xleftmargin=2pt, xrightmargin=2pt,
  linewidth=\columnwidth]
{ "module_name": "pulse_detect",
  "output_signals": [
    { "specs": { "signal_name": "data_out",
                 "direction": "output",
                 "data_type": "wire", "width": 1 },
      "hardware_type": {
        "logic_style": "combinational",
        "clock_domain": null,
        "reset_behavior": null },
      "guarded_commands": { "commands": [
        { "gca_expression": "state == S1 &&
            data_in == 0 : data_out = 1",
          "spec_citation": "The indicating signal
            dataout is 1 at the end cycle of the
            \"pulse\", ..." },
        { "gca_expression": "!(state == S1 &&
            data_in == 0) : data_out = 0",
          "spec_citation": "... returns to 0 until
            the corresponding pulse appears
            again." } ] } } ],
  "internal_signals": [
    "... state: sequential reg, width 2, rising
     clk, async reset to IDLE, 5 commands ..." ],
  "state_transition_table": [
    { "from_state": "IDLE",
      "input_condition": "data_in == 1",
      "to_state": "S1", "output": "data_out = 0" },
    { "from_state": "S1",
      "input_condition": "data_in == 0",
      "to_state": "IDLE",
      "output": "data_out = 1" },
    "... 2 more rows ..." ],
  "fsm_meta": {
    "state_register": "state",
    "encoding": "binary",
    "output_type": "mealy",
    "reset_state": "IDLE",
    "all_states": [ "IDLE", "S1" ] } }
\end{lstlisting}
\caption{Abridged ASTF emitted by the Refinement Module for the
RTLLM \texttt{pulse\_detect} design.
$\mathcal{H}(\texttt{data\_out})$ classifies the indicator as
combinational and \texttt{fsm\_meta} records a Mealy output, so the
emitted RTL asserts \texttt{data\_out} in the pulse's end cycle, the
timing both models miss under direct generation. Ellipses mark
abridged content.}
\label{fig:astf-record}
\end{figure}

\subsubsection{ASTF Auditing}
\label{subsec:audit}
After generating the ASTF, we audit it through five layers before passing it to RTL generation. We use this layered structure because an ASTF must serve as reliable implementation target, and its errors can occur at different refinement level, like unsupported behavior, missing behavior, inconsistent signal logic, and invalid global control structure. 

\begin{itemize}
    \item The first layer checks \textit{soundness} by verifying that each guarded command is grounded in a verbatim citation from the source specification, preventing the LLM from introducing plausible but unsupported behavior. 
    
    \item The second layer checks \textit{completeness} by ensuring that the ASTF covers the required module ports and state behaviors described in the specification, preventing correct but partial representations.
    
    \item The third layer checks \textit{consistency} by validating whether the guarded commands form an interpretable signal-level program, including module-name agreement, declared-signal usage, combinational guard structure, and agreement with truth-table, Karnaugh-map, waveform, or boundary-index evidence when such specification forms are present. 
    
    \item The fourth layer checks \textit{FSM integrity} separately, because FSMs introduce global state constraints that cannot be validated from individual guarded commands alone. Table~\ref{tab:fsm-rules} enumerates the checks, from the declared state skeleton to output timing and framing patterns.
 
    
    \item The fifth layer checks \textit{core RTL design rules}, hardware-legality defects that simulation is weak at catching. Table~\ref{tab:rtl-rules} enumerates each rule, how it is checked on the ASTF, and the defect it prevents. A design can pass every simulation vector yet carry any of these defects into synthesis.

\end{itemize}

\begin{table}[t]
\caption{FSM integrity checks performed by ASTF audit.}
\label{tab:fsm-rules}
\small
\setlength{\tabcolsep}{3pt}
\renewcommand{\arraystretch}{1.08}

\begin{tabularx}{\columnwidth}{
@{}
>{\raggedright\arraybackslash}p{0.23\columnwidth}
Y
>{\raggedright\arraybackslash}p{0.24\columnwidth}
@{}
}
\toprule
\textbf{Check} &
\textbf{What it verifies} &
\textbf{Defect prevented} \\
\midrule

Metadata integrity &
\texttt{fsm\_meta} present and consistent, state register declared,
reset state in the table and state list, encoding width sufficient. &
Undefined or contradictory FSM skeleton. \\

\midrule

Transition completeness &
Every state has outgoing transitions, no unreachable states,
and both branches of binary inputs are covered. &
Dead states and hanging inputs. \\

\midrule

Reset discipline &
Only the reset command references the reset signal. &
Reset logic leaking into state guards. \\

\midrule

Output-type consistency &
Moore outputs read no inputs and agree with the table's outputs. &
Mealy behavior inside a Moore design. \\

\midrule

Output timing &
One-cycle indicators stay Moore, unregistered, and input-free. &
Off-by-one output timing. \\

\midrule

Sequence and framing patterns &
Shift-register compare timing, N-cycle window duration,
and serial start/stop framing. &
Mistimed detection and framing. \\

\bottomrule
\end{tabularx}
\end{table}

\begin{table}[t]
\caption{Core RTL design rules checked by ASTF audit.}
\label{tab:rtl-rules}
\small
\setlength{\tabcolsep}{3pt}
\renewcommand{\arraystretch}{1.08}

\begin{tabularx}{\columnwidth}{
    @{}
    >{\raggedright\arraybackslash}p{0.27\columnwidth}
    Y
    >{\raggedright\arraybackslash}p{0.22\columnwidth}
    @{}
}
\toprule
\textbf{Rule} &
\textbf{How it is checked} &
\textbf{Defect prevented} \\
\midrule

Sequential and combinational logic stay separate &
Enforced by the schema through one
\texttt{logic\_style} per signal; no additional checker is needed. &
Mixed always-block styles. \\

\midrule

Assignment operator matches logic style &
Each \texttt{gca\_expression} is checked against the declared style:
\texttt{<=} for sequential logic and \texttt{=} for combinational
or latch logic. &
Simulation-synthesis mismatch and incorrect hardware classification. \\

\midrule

Combinational logic covers all input cases &
Exact enumeration is used for small input spaces; complement-pair
analysis is used otherwise. &
Unintended latch inference. \\

\midrule

No combinational dependency cycles &
The read-dependency graph among combinational signals is checked
for cycles. &
Electrically unstable feedback. \\

\bottomrule
\end{tabularx}
\end{table}

If any audit layer reports an error, VeriRefine does not pass the ASTF to the Implementation Module. Instead, it converts the detected violation into targeted repair feedback and returns it to the ASTF generation step. The regenerated ASTF is instructed to correct only the reported structural or semantic errors while preserving the remaining valid signal definitions and guarded commands. This refinement-level repair loop allows VeriRefine to correct specification-understanding errors before they are generated into Verilog, where the same errors would be harder to localize from simulation output alone.

\subsection{Implementation Module} 
\label{subsec:implementation}

We use the audited ASTF as the sole input for RTL generation, and implement this validated signal-level representation in synthesizable Verilog.
This design separates implementation from specification interpretation. After the Refinement Module has produced and audited the ASTF, the LLM is no longer asked to infer design intent from the original natural language prompt. Instead, it is constrained to translate explicit signal conditions, state updates, and output assignments into RTL.

Given an audited ASTF, we first identify the RTL generation pattern required by the representation. For combinational designs, the Implementation Module maps ASTF conditions to continuous assignments or combinational always blocks. For sequential designs, it separates next-state or register-update logic from clocked assignments. For FSM-based designs, it preserves the audited state set, transition conditions, and output behavior when constructing the Verilog state machine. In this way, the ASTF serves as the implementation input that determines the structure of the generated RTL.
Code generation is itself rule-governed. The generation prompt enforces structural translation rules, from the port contract to the verbatim translation of every guarded command, and three deterministic checks gate each candidate before simulation, so a violating candidate is regenerated rather than simulated. Table~\ref{tab:codegen-rules} enumerates these rules and the defects they prevent.

\begin{table}[t]
\caption{Coding rules enforced during RTL generation.}
\label{tab:codegen-rules}
\small
\setlength{\tabcolsep}{3pt}
\renewcommand{\arraystretch}{1.08}

\begin{tabularx}{\columnwidth}{
@{}
>{\raggedright\arraybackslash}p{0.23\columnwidth}
Y
>{\raggedright\arraybackslash}p{0.24\columnwidth}
@{}
}
\toprule
\textbf{Rule} &
\textbf{Enforced by} &
\textbf{Defect prevented} \\
\midrule

Port contract &
The module declares exactly the ASTF's ports, names, and widths through prompt guidance and deterministic validation. &
Interface mismatch with the testbench. \\
\midrule

Self-contained module &
The prompt requires that no external submodules are instantiated and every parameter is declared. &
Unresolvable references. \\
\midrule

Style fidelity &
Sequential logic is emitted in clocked always blocks with \texttt{<=}, and combinational logic is emitted as continuous assignments, following $\mathcal{H}(s)$. &
Simulation-synthesis mismatch. \\
\midrule

GCA fidelity &
Every guarded command is translated verbatim in priority order and is never re-derived or optimized. &
Silent logic rewrites. \\
\midrule

Startup definedness &
Registers without a reset receive an \texttt{initial} block. &
X propagation at startup. \\
\midrule

No internal tri-state drivers &
Deterministic checking rejects internal tri-state drivers unless the specification explicitly requires high impedance. &
Conflicting internal drivers. \\
\midrule

Standalone compilability &
The generated module is compiled with \texttt{iverilog} before simulation. &
Syntax and elaboration errors. \\
\bottomrule
\end{tabularx}
\end{table}

\begin{figure}
    \centering
    \includegraphics[width=\linewidth]{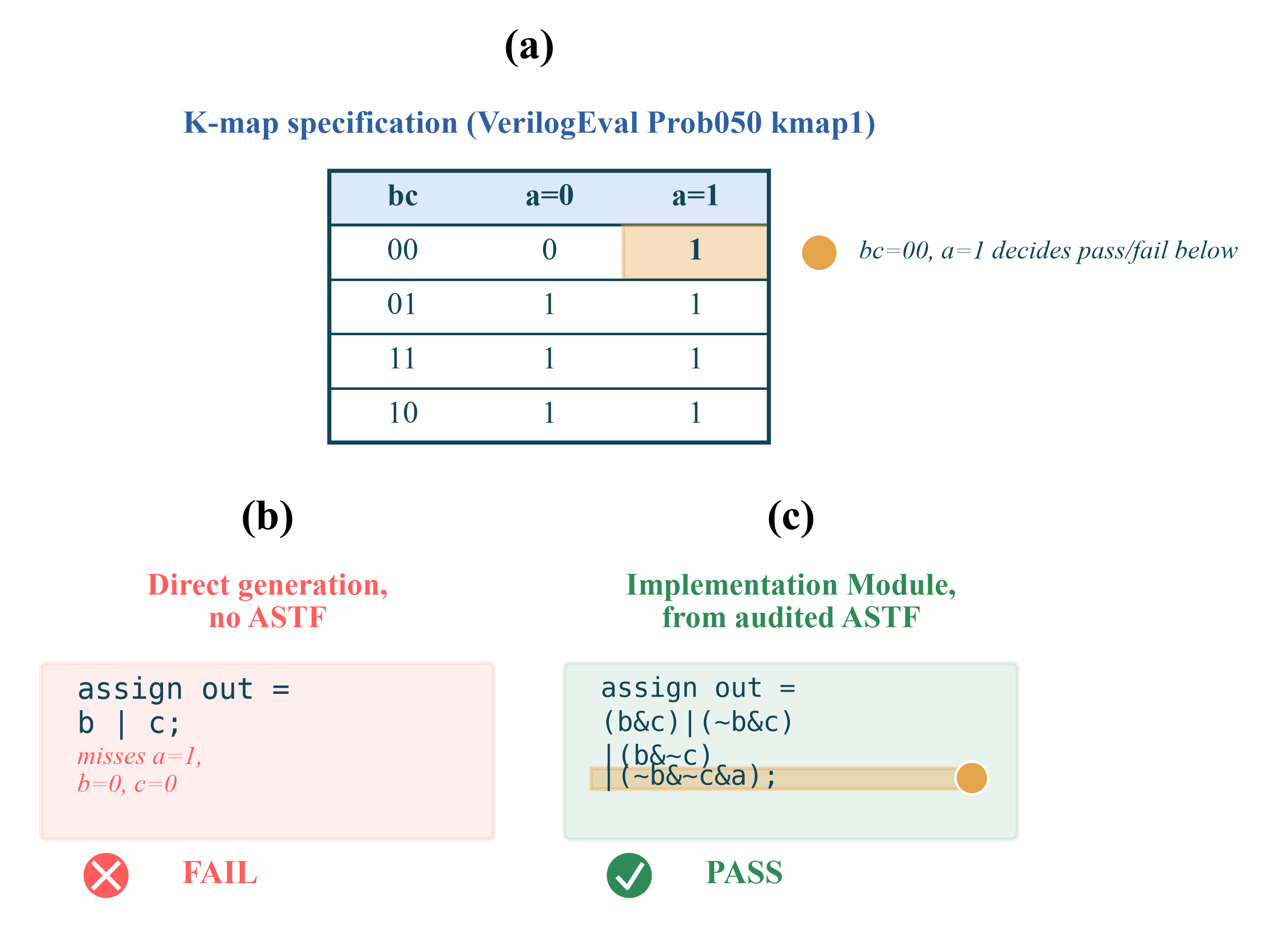} 
    \caption{Prob050 (VerilogEval) example comparing direct RTL generation with VeriRefine.} 

    \label{fig:example}
\end{figure}

Figure~\ref{fig:example} illustrates this ASTF-guided implementation process using the \textit{Prob050 K-map} example. Direct GPT-4-turbo simplifies the K-map to \texttt{b | c}, which misses the asserted minterm \texttt{a \& \textasciitilde b \& \textasciitilde c} and fails simulation. In contrast, VeriRefine first represents the asserted K-map entries in the audited ASTF and then generates RTL from these preserved conditions. The resulting Verilog covers the missing minterm and passes simulation. This example highlights how the Implementation Module makes RTL generation behavior-preserving, structured, and constrained by the audited ASTF.

\subsection{Verification \& Debug Module}
\label{subsec:debug}
\subsubsection{Testbench Generatioin}

The internal testbench (ITB) generator serves as the first functional debug gate in VeriRefine before the generated RTL is evaluated by the official benchmark testbench. ITB is driven only by the natural-language design specification and does not use the generated RTL, the golden reference implementation, the official benchmark testbench, or an externally supplied verification plan during generation. This separation prevents the generated tests from being biased by the candidate RTL. The flow uses a two-stage specification-driven pipeline. First, an LLM converts the specification into a schema-constrained verification plan containing the module purpose, interface signals, clock/reset behavior, assumptions, behaviors to verify, coverage goals, and concrete test scenarios. Each plan contains directed scenarios for functional and corner-case behavior, one LLM-generated explicit 8--15 cycle sequence, and one pseudo-random stress scenario using 80 cycles for combinational designs or 200 cycles for sequential designs.

Second, the plan is converted into a SystemVerilog stimulus module, \texttt{itb\_stimulus}, which drives DUT inputs using a synthetic testbench clock, emits optional \texttt{PROBE} messages, marks scenario completion, and maintains trace metadata such as the current scenario, step index, label, reset state, and packed input snapshots. The LLM-generated stimulus does not instantiate the DUT or implement pass/fail checks. Instead, a deterministic assembly script builds the final simulation harness by connecting the stimulus module, the candidate RTL, and, for VerilogEval, the golden reference RTL. The same clocked harness is used for both combinational and sequential designs; combinational DUTs are not connected to the clock, but their inputs are still sequenced and sampled within the clocked harness. During simulation, the harness applies identical stimulus to the reference and candidate designs and compares their outputs on every negative clock edge after a short settle delay. Mismatches are logged and accumulated rather than terminating the simulation immediately. At the end of the run, the harness emits \texttt{INTERNAL\_TB\_PASS} or \texttt{INTERNAL\_TB\_FAIL}. The simulation log is then summarized into \texttt{sim\_digest.txt}, which reports the verdict, mismatch count, first mismatch time, timeout warnings, completed and failed scenarios, grouped mismatch patterns, and representative failing samples. This digest provides localized debugging evidence for the RTL repair agent while keeping the generated testbench structure reproducible.
\begin{figure}[!t]
    \centering
    \includegraphics[width=\linewidth]{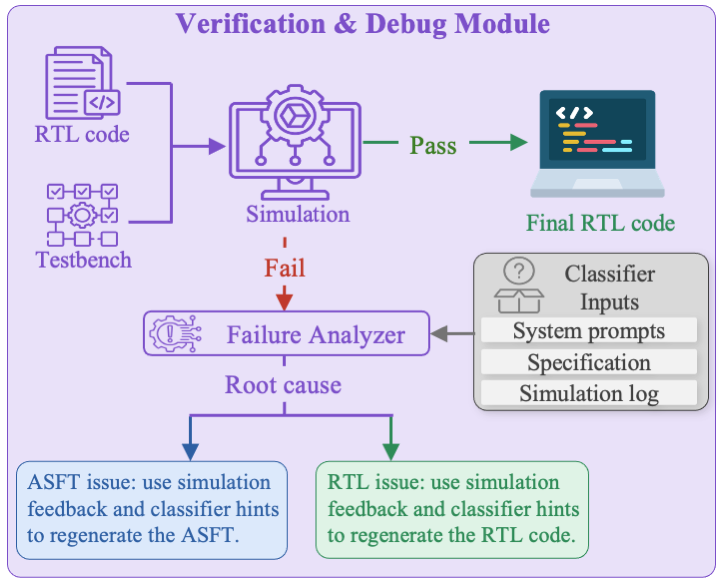} 
    \caption{Verification \& Debug Module}
    \vspace{-8pt}
    \label{fig:debug}
\end{figure}

\subsubsection{Debug}

We design the Verification \& Debug Module to close the loop between generated RTL and executable feedback.
When simulation fails, VeriRefine first determines which module should be repaired. A failure may indicate the ASTF is incorrect, for example because it omits a required signal, encodes an incorrect transition, or assigns an incorrect width. In this case, retrying RTL generation alone cannot fix the design, because the Verilog is generated from a flawed representation. Alternatively, the ASTF is correct, but the generated Verilog implement it incorrectly. In this case, VeriRefine keeps the audited ASTF unchanged and regenerates only the RTL.

To make this decision, the failure analyzer takes three inputs: the original design specification, the current ASTF, and the failure log produced by simulation. The ASTF is summarized into a compact form that includes the module name, signal names, signal widths, logic styles, guarded-command expressions, boundary conditions, and FSM metadata when available. The simulation feedback provides the concrete evidence for the failed attempt, including the pass/fail verdict, the total mismatch count and sample count, the time of the first mismatch, the expected value from the reference module and the actual value from the generated module at each mismatch, and the internal signal states around the mismatch, such as registers, FSM states, and intermediate values. Based on these inputs, the analyzer returns a structured decision that classifies the failure as either an ASTF issue or an RTL issue, together with a concise root cause explanation and actionable repair hints. 

As in Figure~\ref{fig:debug} shown, VeriRefine then uses this decision to route the repair to the appropriate retry loop. If the failure is classified as an ASTF issue, VeriRefine returns to the Refinement Module and regenerates the ASTF with the analyzer's repair hints. If the failure is classified as an RTL issue, VeriRefine keeps the audited ASTF unchanged and retries RTL generation in Implementation Module with the simulation feedback. This two-level routing makes debugging targeted rather than repetitive. The system repairs the representation only when the representation is wrong, and otherwise focuses the repair on the Verilog implementation.

\section{Experiments}
\label{sec:experiments}

\begin{table*}[t]
\centering
\caption{Main results of the four evaluation conditions on \textit{RTLLM v2.0} and \textit{VerilogEval-Human v2}.}
\label{tab:main_results}

\setlength{\tabcolsep}{5pt}
\renewcommand{\arraystretch}{1.2}

\begin{tabular*}{\textwidth}
{@{\extracolsep{\fill}}llcccc@{}}
\toprule

\textbf{Benchmark} &
\textbf{Method} &
\textbf{\makecell{Func.\\Correct. (\%)}} &
\textbf{\makecell{Synth.\\(\%)}} &
\textbf{\makecell{PPA\\Product}} &
\textbf{\makecell{Token\\Cost (M)}}\\

\midrule

\multirow{6}{*}{RTLLM~v2.0}
  & Direct generation (GPT-4-turbo)
  & $25/50~(50.0\%)$
  & $24/25~(96.0\%)$
  & 4.537e-03
  & 0.05 \\

  & Direct generation (Sonnet 4.6)
  & $32/50~(64.0\%)$
  & $30/32~(93.8\%)$
  & 2.761e-03
  & 0.06 \\

  & VeriRefine-Direct (GPT-4-turbo)
  & $34/50~(68.0\%)$
  & $33/34~(97.1\%)$
  & 3.911e-03
  & 1.32 \\

  & VeriRefine-Direct (Sonnet 4.6)
  & $41/50~(82.0\%)$
  & $39/41~(95.1\%)$
  & 4.482e-03
  & 1.53 \\

  & VerilogCoder (Sonnet 4.6)
  & $45/50~(90.0\%)$
  & $43/45~(95.6\%)$
  & 5.046e-03
  & 22.7 \\

  & VeriRefine (Sonnet 4.6)
  & ${47/50~(94.0\%)}$
  & ${45/47~(95.7\%)}$
  & 4.052e-03
  & 3.00 \\

\midrule

\multirow{6}{*}{\makecell{VerilogEval\\Human v2}}
  & Direct generation (GPT-4-turbo)
  & $98/156~(62.8\%)$
  & $97/98~(99.0\%)$
  & 2.366e-04
  & 0.07 \\

  & Direct generation (Sonnet 4.6)
  & $133/156~(85.3\%)$
  & $131/133~(98.5\%)$
  & 4.291e-04
  & 0.09 \\

  & VeriRefine-Direct (GPT-4-turbo)
  & $114/156~(73.1\%)$
  & $113/114~(99.1\%)$
  & 2.040e-04
  & 5.53 \\

  & VeriRefine-Direct (Sonnet 4.6)
  & $143/156~(91.7\%)$
  & $143/143~(100\%)$
  & 2.094e-04
  & 5.91 \\

  & VerilogCoder (Sonnet 4.6)
  & $155/156~(99.4\%)$
  & $153/155~(98.7\%)$
  & 3.300e-04
  & 30.5 \\

  & VeriRefine (Sonnet 4.6)
  & $153/156~(98.1\%)$
  & $151/153~(98.7\%)$
  & 3.321e-04
  & 8.10 \\

\bottomrule
\end{tabular*}
\end{table*}
















\subsection{Experimental Setup}
\label{subsec:setup}
\textbf{Benchmarks.} We evaluate VeriRefine on two established RTL generation benchmarks. \textit{RTLLM v2.0} comprises 50 Verilog design tasks spanning arithmetic circuits, finite state machines, data structures, and control logic, representing complex multi-behavior RTL generation problems. \textit{VerilogEval-Human v2} comprises 156 HDLBits-derived problems covering a broader range of combinational and sequential designs at varying complexity levels. 

\textbf{Models.} We evaluate direct generation and VeriRefine-Direct with two models, GPT-4-turbo and Claude Sonnet 4.6, and run VeriRefine and the VerilogCoder reproduction with Claude Sonnet 4.6 only, all accessed through their official APIs. Generation uses temperature $0$ to minimize sampling randomness, and every evaluation condition shares the same prompts and retry limits.  

\textbf{Evaluation Conditions.} We report four evaluation conditions: 
\begin{enumerate*}[label=(\arabic*)]

\item \textit{Direct generation} (baseline) is a single-pass configuration in which the Verilog is generated directly from the design specification.

\item \textit{VeriRefine-Direct} (ablation) runs the Refinement Module once to produce an audited ASTF and the Implementation Module once to generate Verilog, without invoking the Verification \& Debug Module's repair loop. We detail this configuration in Section~\ref{subsec:ablation}. 

\item \textit{VeriRefine} (our full framework) extends \textit{VeriRefine-Direct} with the Verificaton \& Debug Module. When the generated Verilog fails simulation, the module classifies the failure as an ASTF issue or an RTL issue, and returns the failure feedback to regenerate the ASTF or the RTL accordingly, for a maximum of two repair iterations per design. The complete workflow is described in Section~\ref{sec:method}.

\item \textit{VerilogCoder}~\cite{ho2025verilogcoder} (state-of-the-art baseline) is a published agentic framework for RTL generation. It integrates a planner built on a task and circuit relation graph, a coding agent, and a debug agent that locates faulty signals by tracing simulation waveforms. Its published evaluation reports 94.2\% functional correctness on \textit{VerilogEval-Human v2} with GPT-4-turbo. A direct comparison with VeriRefine requires both frameworks to run on the same model. We reproduced VerilogCoder with Claude Sonnet 4.6 with following the released implementation. It builds on AutoGen~0.2.26 and ships no Anthropic backend, so we implemented a native Anthropic client conforming to AutoGen's \texttt{ModelClient} interface. The client only translates message formats, and all agents, prompts, and per-agent temperature settings remain those of the original release.

\end{enumerate*}

\textbf{Metrics.} We report four metrics: \textit{functional correctness}, \textit{synthesizability}, \textit{PPA quality}, and \textit{token cost}. The first three form a validity hierarchy. A design's synthesizability is only evidence of quality once it passes functional simulation, and its PPA numbers are only evidence
of quality once it synthesizes, so we evaluate each stage only for designs that pass the previous one.

\begin{enumerate}[label=(\arabic*)]
\item \textit{Functional Correctness: }a design is counted as functionally correct when the generated Verilog compiles and passes the benchmark testbench simulation. For each benchmark, we report the number of passing designs and its corresponding percentage. Designs that fail to compile, time out, or produce output mismatches are counted as functional failures.

\item \textit{Synthesizability: }a design is counted as synthesizable when logic synthesis completes without errors and produces a gate-level netlist. We evaluate synthesizability only for functionally correct designs, and report the number that synthesize and the corresponding percentage. This metric separates designs that pass simulation from those a hardware implementation flow can process.

\item \textit{PPA quality} measures the implementation quality of synthesizable designs in terms of power, area, and timing. For each design that passes synthesis, we compute a single PPA product:

\begin{equation*}
\mathrm{PPA} = \left(P_{\mathrm{int}} + P_{\mathrm{sw}} +
P_{\mathrm{leak}}\right) \times A \times T_{\mathrm{cp}}
\label{eq:ppa}
\end{equation*}

where $P_{\mathrm{int}}$, $P_{\mathrm{sw}}$, and $P_{\mathrm{leak}}$ are the internal, switching, and leakage power in $\mu$W reported at the typical corner, $A$ is the die area in $\mathrm{mm}^2$, and $T_{\mathrm{cp}}$ is the critical-path delay in ns. A smaller product indicates a design that is simultaneously lower-power, smaller, and faster.

\item \textit{Token cost} measures the LLM usage a condition requires. We count every input and output token consumed across all designs of the benchmark, passing or failing, including all retries and repair iterations, and report the total in millions. This metric shows costs and allows a fair comparison between conditions on the same benchmark.
\end{enumerate}

\textbf{Implementation.} We implement our framework components in Python. Schema validation uses the \texttt{jsonschema} library with the Draft 2020-12 dialect, and functional simulation uses \textit{Icarus Verilog (iverilog) v~13.0}. 
The synthesizability check runs Yosys 0.30+48 under the open-source SkyWater sky130A PDK (130\,nm), and PPA evaluation runs the OpenLane v1.0.2 flow targeting sky130A with the \texttt{sky130\_fd\_sc\_hd} standard-cell library. The \textit{Verification \& Debug Module} is permitted a maximum of two debug iterations per design. 




\subsection{Key Results and Discussion}





Table~\ref{tab:main_results} reports the four evaluation conditions on both benchmarks under the staged flow of Section~\ref{subsec:setup}. We read the table along the four metrics defined there, from functional correctness to token cost.

\textbf{Functional correctness. }Functional correctness climbs across the evaluation conditions on both benchmarks. On \textit{RTLLM v2.0}, functional correctness climbs across the evaluation conditions for both models. Direct generation passes 25 of 50 designs (50.0\%) with GPT-4-turbo and 32 of 50 designs (64.0\%) with Sonnet 4.6. VeriRefine-Direct lifts the two models to 34 of 50 (68.0\%) and 41 of 50 (82.0\%), an identical gain of 18.0\% that comes from the audited ASTF alone. VeriRefine reaches 47 of 50 (94.0\%) with Sonnet 4.6, and the further 12.0\% come from the verification-guided debug loop.
On \textit{VerilogEval-Human v2}, the same ladder appears from a higher base. Direct generation passes 98 of 156 designs (62.8\%) with GPT-4-turbo and 133 of 156 designs (85.3\%) with Sonnet 4.6. VeriRefine-Direct improves GPT-4-turbo by 10.3\% to 73.1\% and Sonnet 4.6 by 6.4\% to 91.7\%. VeriRefine reaches 153 of 156 (98.1\%) with Sonnet 4.6, adding another 6.4\% through the Verification \& Debug Module. The reproduced VerilogCoder passes 155 of 156 (99.4\%), 5.2\% above its published GPT-4-turbo result.

Comparing the result of these benchmarks reveals a pattern. In absolute numbers, the more designs direct generation fails, the more VeriRefine improves. With Sonnet 4.6, direct generation leaves 36.0\% of designs unsolved on \textit{RTLLM v2.0} but only 14.7\% on \textit{VerilogEval-Human v2}, and VeriRefine correspondingly recovers 30.0\% of the designs on \textit{RTLLM v2.0} against 12.8\% on \textit{VerilogEval-Human v2}. A closer look shows that the two improvements differ in size but not in rate. VeriRefine reduces the number of unsolved designs by 83\% on \textit{RTLLM v2.0} and 87\% on \textit{VerilogEval-Human v2}, nearly the same rate on both benchmarks. This shows that VeriRefine's improvement rate in functional correctness does not depend on benchmark difficulty. At a minimum, the framework's benefit is not limited to one specific benchmark.

\textbf{Synthesizability.} Synthesizability stays within a narrow high band across all six conditions, from 93.8\% to 100\% of functionally correct designs. We take this uniformity as practical confirmation of the metric hierarchy in Section 4.1. Functional correctness is the binding constraint under every condition, and synthesis rarely rejects a design that simulation has accepted. Within the band, the ASTF-constrained conditions hold the upper positions for each model. VeriRefine-Direct with Sonnet 4.6 synthesizes all 143 of its correct designs on \textit{VerilogEval-Human v2}, and VeriRefine passes 45 of 47 on \textit{RTLLM v2.0} against 30 of 32 for direct generation with the same model. We conclude that the designs recovered by refinement and debug are not marginal candidates that fail downstream, and the correctness gains carry through to the netlist stage.

\textbf{PPA quality. }PPA products in Table~\ref{tab:main_results} are averaged over each condition's synthesizable designs, so the rows describe different design sets and are not directly comparable. A condition that solves more designs folds harder designs into its average. Direct generation with Sonnet 4.6 reports the smallest product on RTLLM v2.0 at 2.761e-03, yet this average covers only the 30 designs it synthesizes, the simpler portion of the benchmark. We attribute the small product to this easier design mix rather than to better hardware from direct generation, since the harder designs it fails to solve never enter its average. The informative comparison is therefore between conditions with nearly matched coverage. On \textit{RTLLM v2.0}, VeriRefine synthesizes 45 designs and VerilogCoder 43 from largely overlapping sets, and VeriRefine's product of 4.052e-03 sits about 20\% below VerilogCoder's 5.046e-03. On \textit{VerilogEval-Human v2}, the two frameworks are effectively tied at 3.321e-04 and 3.300e-04. We read these results as evidence that the rule-governed generation of Section~\ref{subsec:implementation} does not sacrifice implementation quality. Fixing each signal's hardware class before code generation leaves power, area, and timing on par with the strongest agentic baseline.

\textbf{Token cost.} 
Token cost matters in deployment because LLM-based frameworks are billed per token and often run under a fixed budget. On \textit{RTLLM v2.0}, direct generation with Sonnet 4.6 spends 0.06M tokens, VeriRefine-Direct 1.53M, and VeriRefine 3.0M. On \textit{VerilogEval-Human v2}, the three conditions spend 0.09M,
5.91M, and 8.1M, and VerilogCoder spends 30.5M. 

VerilogCoder's 30.5M tokens yield 155 correct designs, 197K per correct design, while VeriRefine's 8.2M yield 153, 54K per correct design, a 3.7-fold difference at nearly the same functional correctness rate. Under a fixed token budget, this difference determines how many designs a framework can attempt at all. With a budget equal to VeriRefine's total spend of 8.2M tokens, VerilogCoder's average per-design cost would allow it to attempt roughly 42 of the 156 problems, so even if every attempted design succeeded, it would pass at most 42 against VeriRefine's 153. These results show VeriRefine to be the more token-efficient framework, trailing the state-of-the-art baseline by only two designs (153 against 155) at roughly a quarter of its token cost.

\subsection{Ablation Study}
\label{subsec:ablation}

\begin{figure}[t]
  \centering
  \includegraphics[width=\columnwidth]{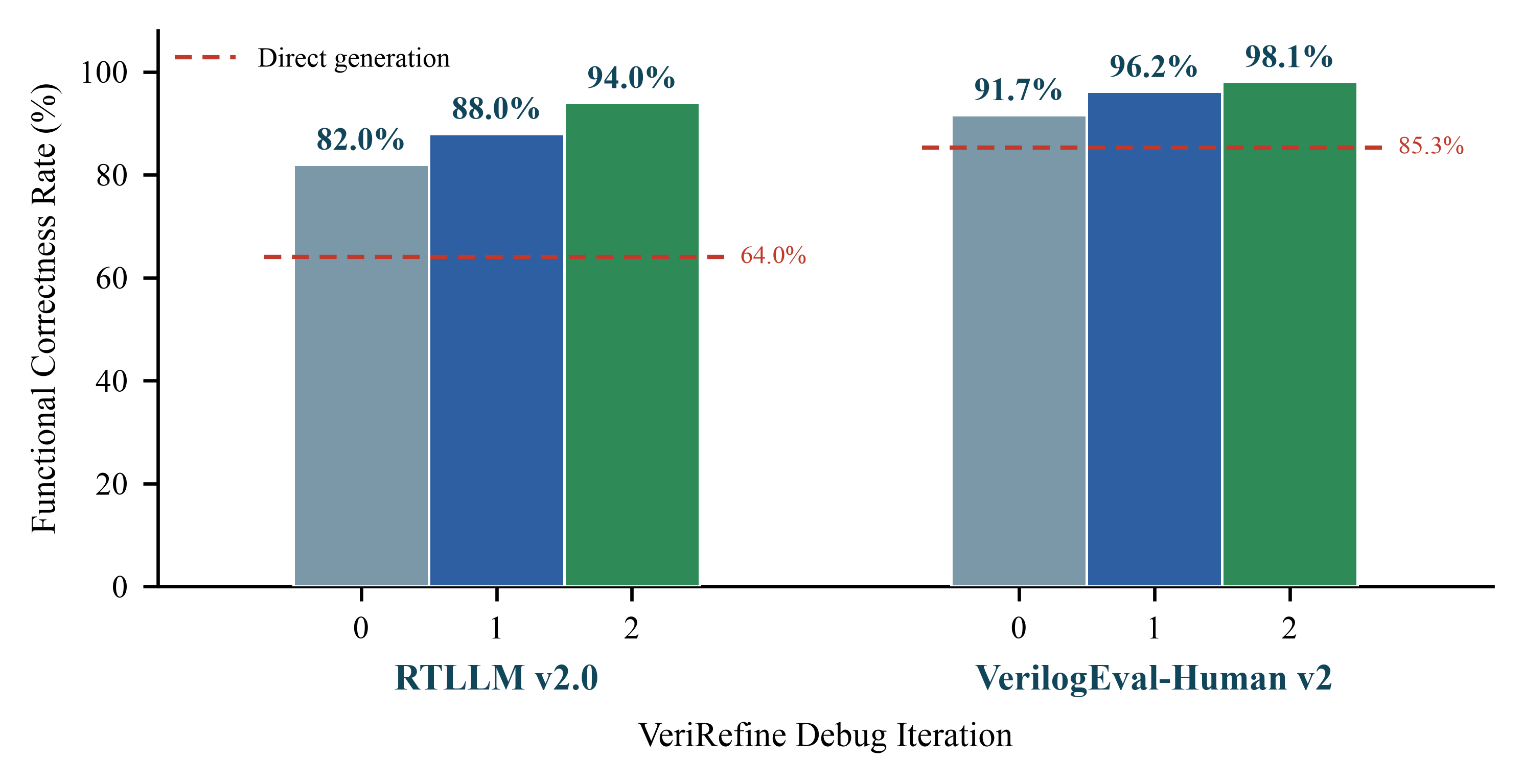}
  \caption{Functional correctness by repair iteration with Sonnet 4.6, where iteration 0 is the single-pass generation corresponding to VeriRefine-Direct and the dashed line marks the direct-generation baseline.}
  \label{fig:ablation}
\end{figure}

We design the ablation study to separate the contribution of the audited ASTF from the debug loop. The ASTF side is measured by VeriRefine-Direct, the single-pass configuration defined in Section~\ref{subsec:setup}, and it appears as debug iteration 0 in Figure~\ref{fig:ablation}. The debug side is measured by the two iterations that follow, the maximum our setting allows. In iteration 1, the Verification \& Debug Module receives a failure log of the design that failed simulation, classifies failure as an ASTF issue or an RTL issue, and regenerates the ASTF or the RTL with that feedback before re-simulating. Iteration 2 repeats the same procedure for the designs whose first fix still fails, this time guided by the failure log of the repaired candidate rather than the original one. 
Figure~\ref{fig:ablation} reports the outcome of this
process. On \textit{RTLLM v2.0}, iteration 1 recovers 3 designs and lifts functional correctness from 82.0\% to 88.0\%, and iteration 2 recovers 3 more to reach 94.0\%. On \textit{VerilogEval-Human v2}, the two iterations recover 7 and 3 designs, lifting 91.7\% to 96.2\% and then 98.1\%.

Through this analysis, we find that the audited ASTF is the primary contributor to VeriRefine's improvement. It contributes 18.0\% of the total 30.0\% on \textit{RTLLM v2.0} and 6.4\% of 12.8\%
on \textit{VerilogEval-Human v2}, and already places iteration 0
well above the direct-generation baselines of 64.0\% and 85.3\%.
The debug iterations contribute the rest by converting each
concrete failure into a targeted fix. We therefore conclude that
VeriRefine's strongest ability lies in the first-pass generation
the audited ASTF enables, with the debug loop completing the
remainder.

\subsection{Failure Cases Analysis}

VeriRefine has six designs unsolved across the two benchmarks, three on \textit{RTLLM v2.0} and three on \textit{VerilogEval-Human v2}. We investigate each failure and group their root causes into three categories.

We attribute three of the failures, \texttt{radix2\_div}, \texttt{freq\_divbyeven}, and \texttt{Prob099}, to benchmark defects, where the specification and the benchmark testbench contradict each other, so no implementation faithful to the specification can pass. \texttt{radix2\_div} fails because the testbench requires a \texttt{res\_ready} handshake that the specification never mentions. \texttt{Prob099} fails because the testbench connects outputs \texttt{Y2} and \texttt{Y4} to a reference module that does not declare them, so simulation stops before any behavior is tested. \texttt{freq\_divbyeven} carries the same class of specification-to-testbench mismatch.\linebreak \texttt{freq\_divbyeven} fails because the specification names the module \texttt{freq\_diveven} while the testbench instantiates \texttt{freq\_divbyeven}, and the testbench further assumes a division factor of six that the specification leaves unspecified. Our internal testbench, generated from the specification alone, passes the \texttt{radix2\_div} and \texttt{Prob099} designs that the official testbench rejects, which locates the defect between the specification and the testbench rather than in the generated RTL.

The \texttt{serial2parallel} failure stands apart as specification ambiguity. Its specification allows the output valid signal to be asserted either in the cycle of the eighth input bit or one cycle later, and the testbench accepts only the later reading. Our audit checks whether the ASTF is faithful to the specification, and both readings are faithful, so this failure class is invisible to the audit by design.

The remaining two failures, \texttt{Prob149} and \texttt{Prob153}, reflect capability limits. The two designs, a sequential control design and a gshare branch predictor, fail with a mismatch signature that stays nearly constant across all regenerated candidates. A constant signature points to a stable misreading of the required logic rather than random error, and more debug iterations alone would not close it.



\section{Conclusion}
\label{sec:dis}

We presented VeriRefine, an agentic framework that separates specification understanding from code generation through the ASTF, a per-signal intermediate representation with verbatim provenance. A closed schema and a five-layer audit verify each ASTF before any
Verilog exists, and a spec-only internal testbench drives the debug loop after simulation. With Claude Sonnet 4.6, VeriRefine reaches 94.0\% functional correctness on \textit{RTLLM v2.0} and 98.1\% on \textit{VerilogEval-Human v2}, within two designs of the state-of-the-art baseline at roughly a quarter of its token cost with synthesizability and PPA quality on par with that baseline, and the ablation shows the audited ASTF is the primary contributor.

The failure analysis also marks the boundary of VeriRefine and points to our future work. A framework fully faithful to its specification inherits the specification's defects, and four of the six remaining failures fall exactly there, where the specification contradicts the official testbench or leaves a behavior open to more than one understanding. Our future work therefore targets specification refinement. Instead of trusting the specification unconditionally, the
framework should question it, checking the specification's internal consistency and flagging contradictory sentences, underdetermined behaviors, and forced assumptions before they propagate into the ASTF. The framework would then improve not only the RTL generation but also the quality of the specification itself, making it a more
correct and precise description of the intended design.

\section*{Acknowledgment}




This work is supported by the National Science Foundation under Grant No. 2434247. Any opinions, findings, and conclusions or recommendations expressed in this material are those of the author(s) and do not necessarily reflect the views of the funding agencies.

\bibliographystyle{ACM-Reference-Format}
\bibliography{refs}

\end{document}